\documentclass[preprint,12pt]{elsarticle}

\usepackage{amssymb}

\usepackage[nodots]{numcompress}

\usepackage{graphicx,epstopdf}
\usepackage{txfonts,enumerate}
\usepackage[hyphens]{url}

\journal{JQSRT}

\def\magn{mag$_{\rm SQM}$ arcsec$^{-2}$}
\def\cdm{cd m$^{-2}$}
\def\mcdm{mcd m$^{-2}$}
\def\scotopiclimitmag{18.9 mag$_{\rm SQM}$ arcsec$^{-2}$}
\def\scotopiclimitcdm{$3 \times 10^{-3}$ cd/m$^2$}

\begin{document}

\begin{frontmatter}



\title{Night sky photometry and spectroscopy\\
performed at the Vienna University Observatory}


\author{Johannes Puschnig, Thomas Posch, Stefan Uttenthaler}

\address{Universit\"at Wien, Institut f\"ur Astrophysik, 
T\"urkenschanzstra{\ss}e 17, A-1180 Wien, Austria}

\begin{abstract}
We present night sky brightness measurements performed at the Vienna University Observatory and at the Leopold-Figl-Observa\-torium f\"ur Astrophysik, which is located about 35km to the southwest of Vienna. The measurements have been performed with Sky Quality Meters made by Unihedron. They cover a time span of roughly one year and have been carried out every night, yielding a night sky brightness value every 7 seconds and thus delivering a large amount of data. In this paper, the level of light pollution in Vienna, which ranges from 15 to 
19.25 \magn\ is presented for the very first time in a systematic way.
We discuss the influence of different environmental conditions on the night sky brightness and implications for human vision. We show that the circalunar rhythm of night sky brightness is extinguished at our observatory due to light pollution.

Additionally, we present spectra of the night sky in Vienna, taken with a 
0.8m telescope. The goal of these spectroscopic measurements was to identify the main types of light sources and the spectral lines which cause the light pollution in Vienna. It turned out that fluorescent lamps are responsible for the strongest lines of the night sky above Vienna (e.g.\ lines at 546\,nm 
and at 611\,nm).
\end{abstract}


\end{frontmatter}


\section{Introduction \label{sec:intro}}

Up to now, astronomers have mainly been measuring the brightness of the night sky (in magnitudes/arcsec$^2$) at dark sites, especially at modern mountain observatories, or at potential observatory sites as a part of ``site-testing'' and ``site-monitoring''. Within the past few years, it has become evident that increasing night sky brightness and light pollution have far-reaching consequences for many branches of human life as well as wildlife (e.g.\ Posch et al. \cite{Posch2010}).

Therefore, it is desirable to measure and monitor the night sky brightness not only at remote mountaintop observatory locations (as done, e.g., by Patat \cite{Patat2008}), but also close to the centers of modern civilization, and to do so every night, in a reproducible way, with the aim of performing long-term studies (such as in climate research). Only in this way can the impact of night sky brightness on biological rhythms on animal behaviour, and human health be assessed.

This is the aim of our ongoing night sky brightness measurements in Vienna and at the ``Leopold-Figl-Observatorium f\"ur Astrophysik'' (LFOA) on Mount Mittersch\"opfl. We complemented our measurements with spectroscopic studies of the night sky at the Vienna University Observatory, which hosts the ``Institut 
f\"ur Astrophysik'' (IfA).

It should be noted that all our measurements refer to scattered light. We measured only brightness values and spectra of the sky
background, while we explicitly avoided to measure any direct radiation
from streetlamps. This is what we call ``night sky brightness'' (NSB).
In other words: the brightness values and the spectra presented in this
paper refer to the total backscattered light of the night sky. Its origin
is the whole ensemble of streetlamps, facade illuminations, illuminated
billboards etc.\ at the respective observing site and in its near and far surroundings. The natural nocturnal radiation from the Earth's atmosphere, which is produced by different processes such as recombination of atoms that have been ionized by the Sun's radiation during daytime, contributes very little to the NSB that we measured at our observing sites, since the latter is dominated by the influence of artificial light.


\section{Measurement sites and methods \label{sec:methods}}

\subsection{Measurement sites}

First measurements began in November 2011 at the IfA. Since April 2nd, 2012 the measuring device (label ``IFA'') is mounted at its ultimate place and points exactly to the zenith. About the same time (March 21st, 2012) we started measuring the NSB at our remote mountaintop site too. Currently we use three devices in total, two of which are located at LFOA (label ``FOA'' and ``FOA2''). The geographical positions of our sites are given below.

\begin{table}[h]
\caption{Geographical coordinates of our measurement sites. The distances of the sites IfA and LFOA from the city center are 3.5\,km and 35\,km, respectively.}
\begin{center}
\begin{tabular}{| c | c | c |}
  \hline
	Location (altitude) & Longitude & Latitude \\ \hline
	IfA (240m)  & 16$^{\circ}$ 20' 03' E  & 48$^{\circ}$ 13' 54'' N \\
	LFOA (880m) & 15$^{\circ}$ 55' 24'' E & 48$^{\circ}$ 05' 03'' N \\ 
  \hline
\end{tabular}
\end{center}
\label{t:sites}
\end{table}


\subsection{Photometric measurements}

\subsubsection{Units used for our photometric measurements}

The devices that we use for our NSB measurements yield data in a unit
which is very widespread in astronomy, namely magnitudes per square
arcsecond (mag/ arcsec$^2$). This unit corresponds to a luminance, but is a logarithmic measure derived from stellar photometry, where larger
values correspond to fainter objects. In the same way, larger NSB values
in mag/arcsec$^2$ indicate darker skies (with less light pollution).
Equation (\ref{eq:unihedron}) gives the conversion from mag\,arcsec$^{-2}$ to \cdm\ and Table \ref{t:conversion} lists selected pairs of
corresponding values.

\begin{equation}[h]
   Luminance\ [cd/m^2] = 10.8\times10^4 \times 10^{(-0.4\ \times\ [mag arcsec^{-2}])}
   \label{eq:unihedron}
\end{equation}

\begin{table}[h]
\caption{Conversion between mag arcsec$^{-2}$ and \mcdm.}
\begin{center}
\begin{tabular}{| l | r| l |}
  \hline
mag/arcsec$^2$ & \mcdm\ & comment \\ \hline
14  &  271 &  \\
15  &  108 &  \\
16  &  43.0 &  \\
17  &  17.1 &  \\
18  &  6.81 &  \\
19  &  2.71 &  \\
20  &  1.08 &  \\
21  &  0.430 &  \\
21.75  &  0.215 &  value we assume for the {\em overall} \\
 & &  natural clear sky brightness \\
22.0   &  0.172  &  value we assume for the {\em zenithal} \\
 & &  clear sky brightness \\
\hline
\end{tabular}
\end{center}
\label{t:conversion}
\end{table}

For the purpose of monitoring the night sky brightness over a long period of time, we use Unihedron's Sky Quality Meter with an integrated lensing system. The model is called ``SQM-LE'', but hereafter we will refer to it as ``SQM''. All our devices are connected via ethernet. According to the manufacturer's manual, the integrated lens narrows down the field of view to only $\approx$20$^{\circ}$ full width at half maximum (FWHM), thus at an off-axis distance of $\approx$10$^{\circ}$, the sensitivity declines by a factor of 2. At higher angles the response decreases rapidly, so that a point source located $\approx$19$^{\circ}$ off-axis contributes a factor of 10 less to the measured brightness than on-axis. We have chosen the lensed version because we also use the devices in urban regions where the field of view is limited due to surrounding buildings. A wide beam width then could result in lower accuracy due to the possible influence of nearby lights.

An advantage of the ethernet version is that with network cables far greater distances can be reached than with USB cables and by making use of the ``Power over Ethernet'' (PoE) technology also power supply at remote sites is easier to handle. Furthermore, the PoE device directly attached to the SQM leads to a sufficient amount of heating as to avoid the formation of dew.

\subsubsection{Spectral response function of the SQM and comparison to the Johnson V band}

Sky Quality Meters are equipped with a photo diode, a filter and a temperature sensor for thermal stabilization. The manufacturer provides information on the sensitivity curve of the photo diode and the filter, but not the resulting response function.
Previous reports indicated that the night sky brightness values measured by SQMs do not differ strongly from those from systems equipped with a Johnson V filter. Nevertheless, given that SQM magnitudes are not the same as V magnitudes, we will use the unit ``\magn'' throughout this paper in order to make this difference clear. 

The spectral sensitivity of the SQM was measured by Cinzano 
\cite{Cinzano2005}. A comparison between the SQM response function and the Johnson V band shows that the SQM is much more sensitive to light below $\approx$520\,nm. At a first glance the difference seems to make a calibration between the Johnson V band and the SQM response function difficult, but considering the spectral distribution of the night sky over Vienna which is dominated by emission lines above $\approx$525\,nm as seen in Fig.\ \ref{fig:skyspectrum}, we presume that a constant offset can be found for a comparison to the Johnson V band.

Measurements by other groups indicate that the SQM actually underestimates
the brightness of the night sky compared to V band measurements -- precisely
due to its enhanced sensitivity in the blue (which is less prominent in the night sky than e.g.\ in the twilight sky). This is supported also by our own
preliminary comparative measurements with instruments equipped with a 
Johnson V filter. The difference may amount up to 0.6\,mag.\
for clouded skies, while it is smaller for clear skies (H.\ Spoelstra, 
priv.\ comm.).

\begin{figure}[t]
	\centering
	\includegraphics[width=13cm]{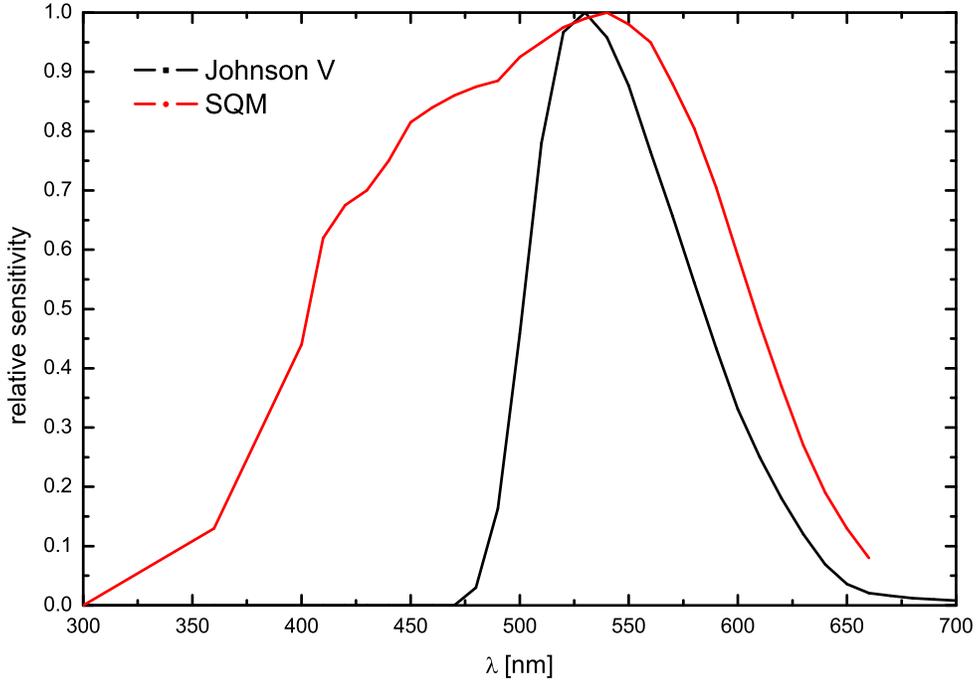}
	\caption{SQM Response function as measured by Cinzano \cite{Cinzano2005} in comparison to the Johnson V filter response function defined by the Johns Hopkins University \cite{Dobos2004}.}
	\label{fig:sqmresponse}
\end{figure}

For the moment both SQM and Johnson V measurements seem to be suitable to quantify artificial NSB. But having in mind that light emitting diodes (LEDs) are developing fast, a change in the night sky brightness composition can be expected. Since future LEDs of color temperatures above 3000K may have a significant emittance in the blue part of the spectrum, Johnson V band measurements would not be able to register a change in color of the NSB. Since the SQM is sensitive to the blue end of the spectrum, it seems to be prepared for future measurements of the sky brightness in case of stronger contributions
at the blue end of the spectrum.
The best one can do without spectroscopy is attaching a set of SQMs parallel to each other, each equipped with a standard photometric filter (LRGB or UBVR).
Such a configuration makes it possible to study the NSB development and to
record possible changes in the color of the night sky at the same time (see Kyba et al. \cite{Kyba2012a}).

\subsubsection{Sampling rate}

We started our measurements with a sampling of 5 minutes, but soon it was noticed that with overcast skies, we miss some features within our light curves. Therefore in the next step the frequency was increased to a reading once per minute. This seems to be a good value for most applications and a good compromise between data storage, data handling and valuable information. But since we want to measure all possible variations that can occur, we even increased the sampling rate to a reading every 7 seconds which corresponds to a frequency of 0.143 Hz. The device measures accurately at that high frequency and so far we did not encounter any problems due to integration time limitations. We are able to detect lightning, fireworks and other short-term events. Moreover, with higher sampling it is possible to quantify the influence of the curfew in Vienna even in cloudy nights, when the light curves are much more disturbed.

\subsubsection{Automated data storage and publishing}

All our data are logged and stored automatically at each monitoring station. From there every five minutes a synchronization script uploads the data to the University's web server for data mirroring and data publishing leading to ``real-time'' data visualization on our website\\
\url{http://astro.univie.ac.at/en/institute/light-pollution/}. 

A tool is provided for calculating all relevant information about illuminated moon fraction, moon rise, moon set and dates for civil, nautical and astronomical twilights. Furthermore we calculate mean, best and worst values as well as statistical information for the defined time intervals (e.g.\ nautical twilight). Some other values like the ``naked eye limiting magnitude'' or the ``clear sky factor'' are calculated, but still in an experimental phase. A future discussion on that is planned.

\subsubsection{Intercomparison and relative metering precision of our two SQMs}

For checking the quality of the photometric data, we have made a parallel measurement with our two devices ``IFA'' and ``FOA2'' over a period of 33 days. Since an absolute calibration to any given standard such as Johnson V seems to be difficult for the moment, the question to be answered was, weather both of our devices give similar response under the same environmental conditions. Therefore, the devices were installed at exactly the same place at our institute and pointed to the zenithal region of the sky.
The daily comparative light curves observed can be found in the appendix (Figs.\ \ref{fig:qualitylightcurves1} to \ref{fig:qualitylightcurves3}).

Analysis of the data showed that for the substantial fraction of measurements the differences of both device data readings lie within an interval of only $\approx$0.1 \magn. But it was also recognized that under certain circumstances scattered moonlight caused differences of up to $\approx$0.9\magn. A discussion on that topic can be found in the Appendix.

\begin{figure}[t]
	\centering
	\includegraphics[width=13cm]{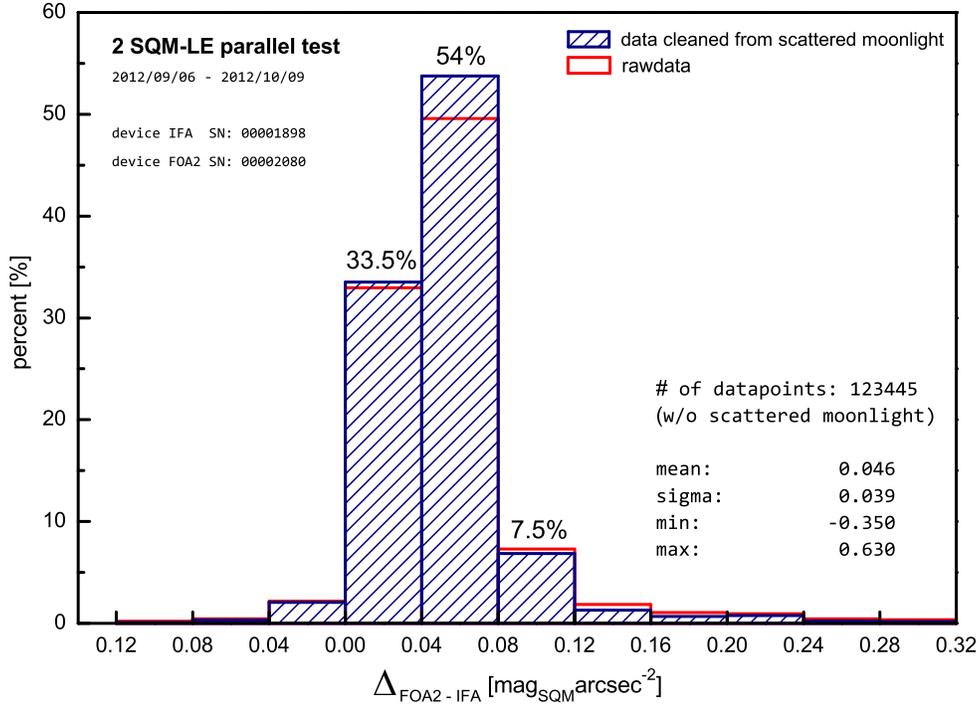}
	\caption{Histogram showing the distribution of the differences between the measurements of 2 SQMs operated in parallel mode. Red lines mark distribution before correction for scattered moonlight. The bin size is 0.04 \magn.}
	\label{fig:qualityhistogram}
\end{figure}

Neglecting scattered moonlight values, the results show that the initial calibration of our SQMs is offset by only $\approx$0.046 \magn. Taking that offset into account the relative accuracy between the two devices is given by the standard deviation of all data points. Fig.\ \ref{fig:qualityhistogram} shows the distribution of the calculated differences ($\Delta$) binned to 0.04 \magn. It can be seen that after rejecting data with scattered moonlight for 87.5 \% of all data points the difference in the response of our devices is equal to or lower than 0.08 \magn\ leading to a standard deviation of $\approx$0.04 \magn\ which corresponds to a relative error of only 4 \%. Under the assumption of a 3 \% growth of the NSB for Vienna, we will be able to detect a significant ($2\sigma$) change of the NSB level after 3 years (by the end of 2015).

So far the SQM appears to be a handy and robust instrument for long term studies of the development of the nightsky brightness. Under some circumstances scattered moonlight can influence the readings, therefore some sort of shielding should be used for the lensed SQM  version.

\subsection{Spectroscopic measurements}

The spectroscopic observations, presented below, were done with a 80\,cm Cassegrain telescope in the North Dome of the Vienna University Observatory. Its focal length amounts to 6.7\,m. A spectrograph 'DSS-7' produced by SBIG was attached to the telescope. DSS-7 is a grating spectrograph offering five slits with different widths, the narrowest of which provides for a dispersion of $\sim0.55$\,nm/pixel, corresponding to a spectral resolving power of $\lambda$/$\Delta$$\lambda$ $\approx$ 500. An ST-7 CCD camera, also made by SBIG, was connected to the spectrograph to acquire the spectra. With this camera, the wavelength coverage is approximately 450--850\,nm (with some small variations between the observing runs). The chip of the CCD camera was cooled down to -10$^{\circ}$\,C during the observations. 

To record the spectra of the night sky, the telescope was pointed to the south or to the south-east (i.e.\ towards the city center of Vienna) at an elevation of $\sim$45$^{\circ}$. Frames of 300\,s integration time were obtained. All spectroscopic observations were carried out during the evening hours. Some of our night sky spectra were obtained as dedicated observations, others were extracted as a by-product of stellar spectroscopy, in which case they were subtracted as background. The 2D raw spectra were dark-subtracted and flat-fielded with dome flats (using light bulbs mounted on the telescope). To extract the 1D spectra, all pixel rows that were illuminated by the narrowest slit were summed up along the slit direction. The wavelength calibration was done using the sky emission lines themselves by adopting a second-order polynomial to map pixel positions to wavelength values.

We expect that night sky spectra will change worldwide during the next few decades due to increasing numbers of light emitting diodes (LEDs) used for street and facade lighting. It is desirable to monitor the resulting changes in the average night sky spectra in order to see whether indeed the relative flux at short wavelengths ($\lambda$ $<$ 500\,nm) will thereby increase (which would be unfortunate for many reasons).


\section{Data analysis}

In the following all given daily mean values in \magn\ and other statistical parameters for the NSB are calculated for time intervals between the astronomical twilights of the given date. As described in Section~\ref{sec:methods} the device ``FOA2'' mounted at our remote site was calibrated against our older SQM ``IFA'' located at our institute in Vienna.

\subsection{Detection of the natural circalunar rhythm}

\begin{figure}
	\centering
	\includegraphics[width=11.5cm]{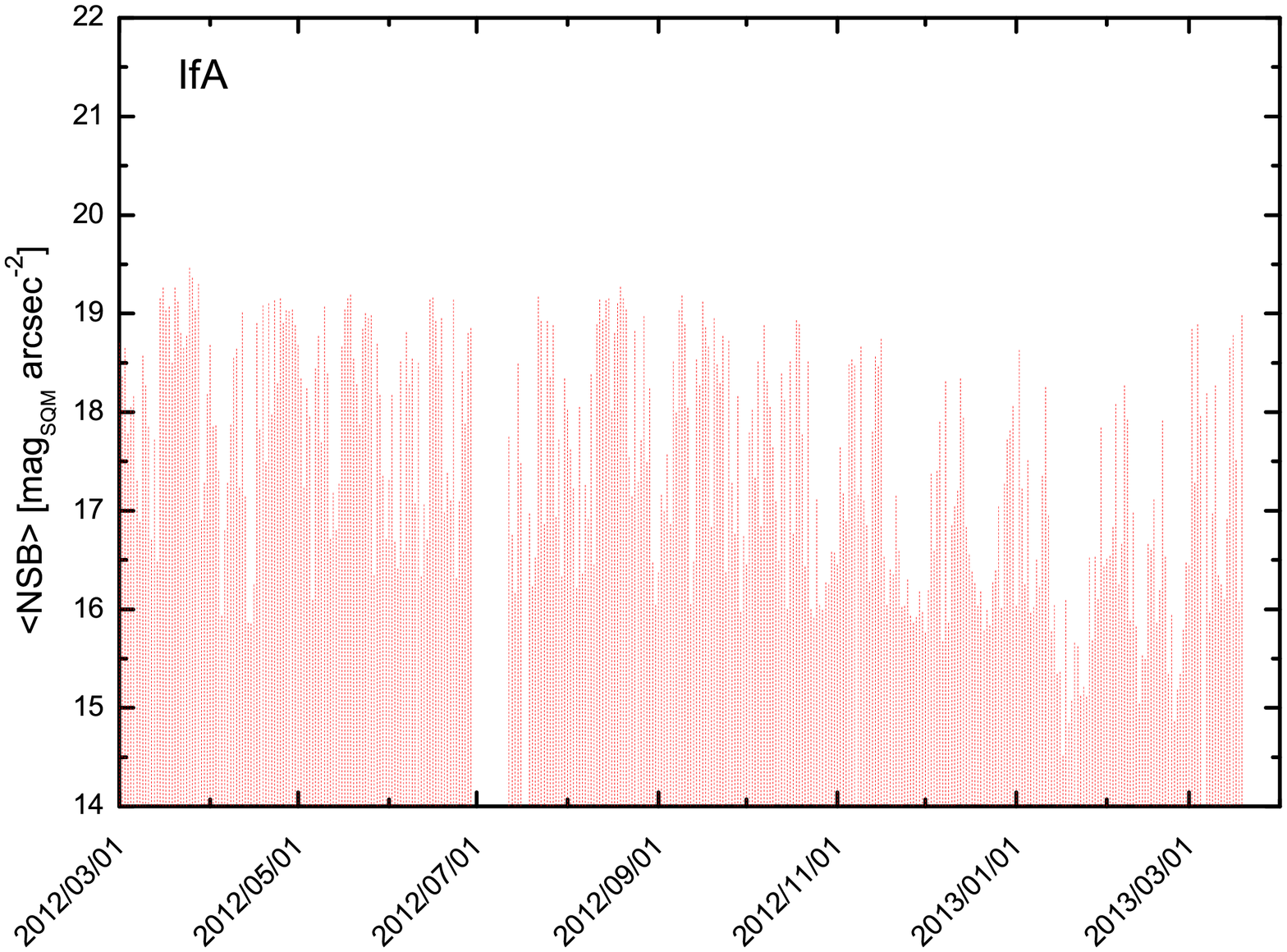}
	\includegraphics[width=11.5cm]{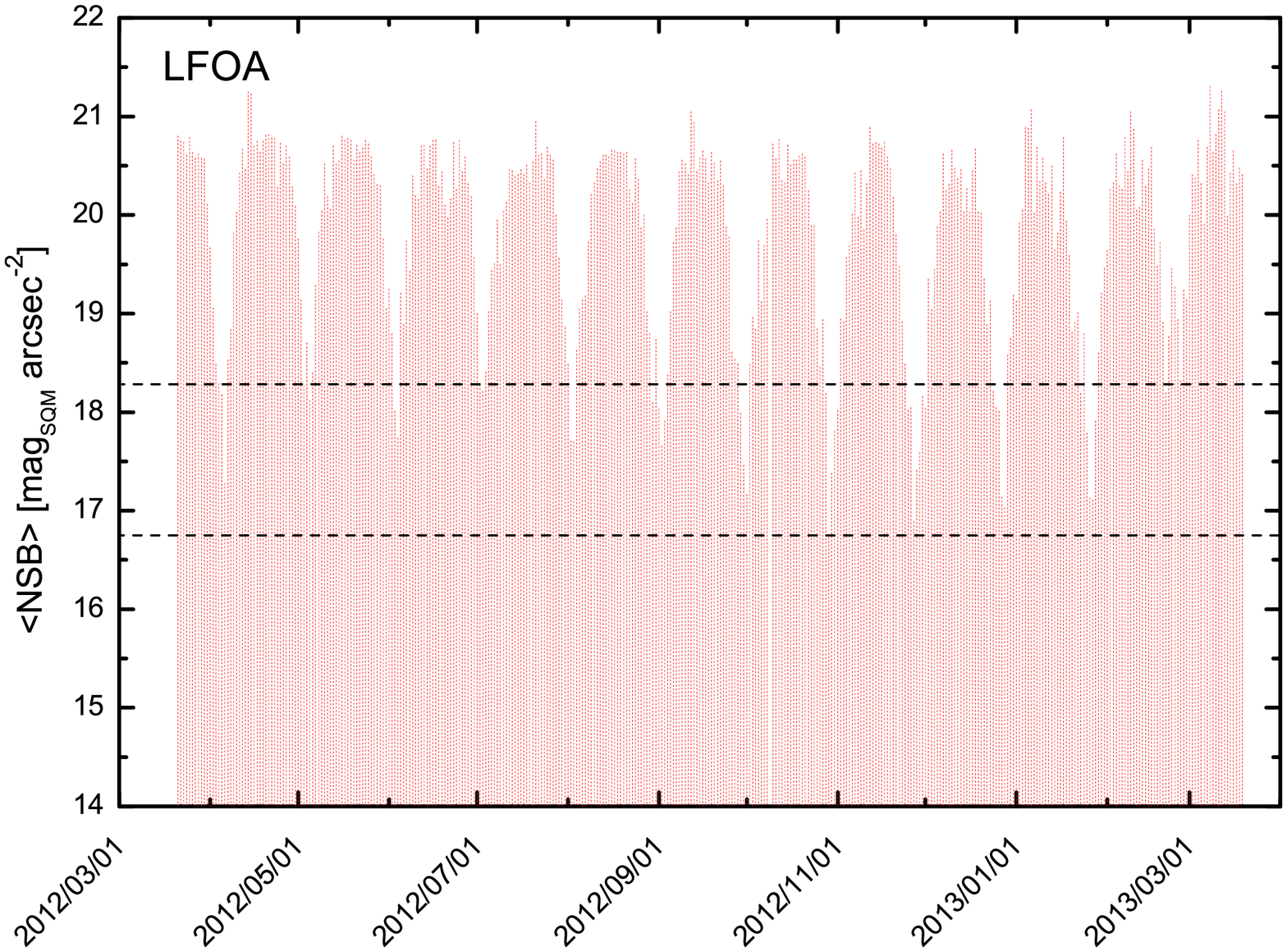}
	\caption{Mean daily NSB within astronomical twilight. Distribution over one year (date format: yyyy/mm/dd) obtained at both sites IfA (top) and LFOA (bottom). At LFOA the circalunar rhythm dominates the variations, whereas in Vienna the natural variation is barely recognizable. The LFOA data also show a circannual period depending on the maximum elevation of the full moon (see dashed lines).}
	\label{fig:lightcurve}
\end{figure}

Our measurements meanwhile cover a period of one year. We were therefore interested in the behavior of the mean night sky brightness at IfA and at LFOA during this period (March 2012 -- March 2013).
At a rural site, a strong influence of the lunar phase on the NSB is expected: full moon nights should be recognizable by much brighter night skies, especially for clear sky conditions. In contrast, nights around new moon should be much darker. Since LFOA is a rural site, this pattern is indeed detected and a nice circalunar periodicity of the mean NSB becomes evident (see Fig.\ \ref{fig:lightcurve}).
The NSB data from Vienna show a completely different picture. Instead of the moon, the degree of cloudiness has the strongest influence on the NSB in Vienna. The more clouds over a big city, the stronger the backscattering of urban light. Of course, the degree of cloudiness does not follow any strict periodicity. Therefore, the mean NSB measured in Vienna seems to vary in a stochastic way. The circalunar rhythm is barely recognizable here.

The lower half of Fig.\ \ref{fig:lightcurve} also shows a circannual periodicity of the NSB close to full moon: winter nights with the full moon at higher elevations are significantly brighter (by $\approx$ 1.5\,\magn) than summer nights.

\subsection{Range of NSB values under urban conditions}

\begin{figure}
	\centering
	\includegraphics[width=11.5cm]{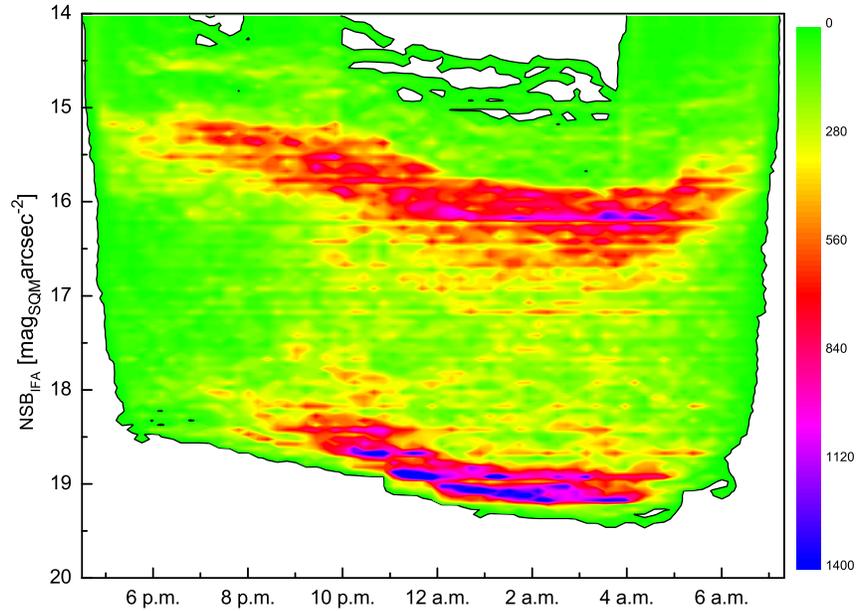}

	\caption{Density plot of all measurements obtained at Vienna (IfA) from April 2012 to April 2013. A gradient of 0.1 \magn\ per hour and two dominant NSB ranges can be identified.}
	\label{fig:IFAdarkness}
\end{figure}

Figures \ref{fig:lightcurve} (upper part) and \ref{fig:IFAdarkness} clearly show that the NSB values we measure at our urban site IfA vary within a large range. The darkest skies reached at IfA are characterized by an NSB slightly above (= darker than) 19.25 \magn, corresponding to a zenithal luminance of 2.15 \mcdm. The brightest skies reached at IfA under overcast conditions have an NSB below (= brighter than) 15 \magn, corresponding to a zenithal luminance of more than 100 \mcdm, which is more than 0.1 \cdm\ and can easily be measured even with a simple luminance meter.
As Fig.\ \ref{fig:IFAdarkness} shows, not all NSB values occur equally often. The contour plot -- which is based on more than 2 $\times$ 10$^{6}$ individual data points -- shows two values which occur much more often than others: one is 19.1 \magn\ -- with a trend to still slightly larger (darker) values as the early morning hours (with less traffic and less artificial lights) are reached. These values are characteristic for clear skies (no clouds, little aerosol content of the air, relatively little amount of backscattering). The other range of frequently occurring NSBs is about 16.3\,\magn, corresponding to 33\,\mcdm. This value is about 13 times larger than the former one and corresponds to completely overcast conditions. The values between 16 and 19\,\magn\ are measured when the sky is partly cloudy, or cloudless but hazy, or clear but not moonless.
Furthermore Fig.\ \ref{fig:IFAdarkness} shows that the trend of NSB development in Vienna during one night typically follows a gradient of 0.1 \magn\ per hour.

It is also interesting to compare our ``best'' measured values for the urban conditions at IfA with the predited sky brightness that we extracted from Cinzano, Falchi \& Elvidge \cite{Cinzano2001} in their ``World Atlas of the artificial night sky brightness''. We find that our urban observatory IfA has a predicted clear-sky NSB of approximately 10 times the natural night sky brightness. Our above reported best value of 19.25\,\magn\ or 2.15 \mcdm\ amounts to 12.5 times the natural zenithal night sky brightness, for which we assume 0.172\mcdm\ (Duriscoe, priv. comm.; see also Tab. \ref{t:conversion}).

\subsection{Detecting the curfew(s) in Vienna (11 p.m., 12 p.m.)}

\begin{figure}
	\centering
	\includegraphics[width=12cm]{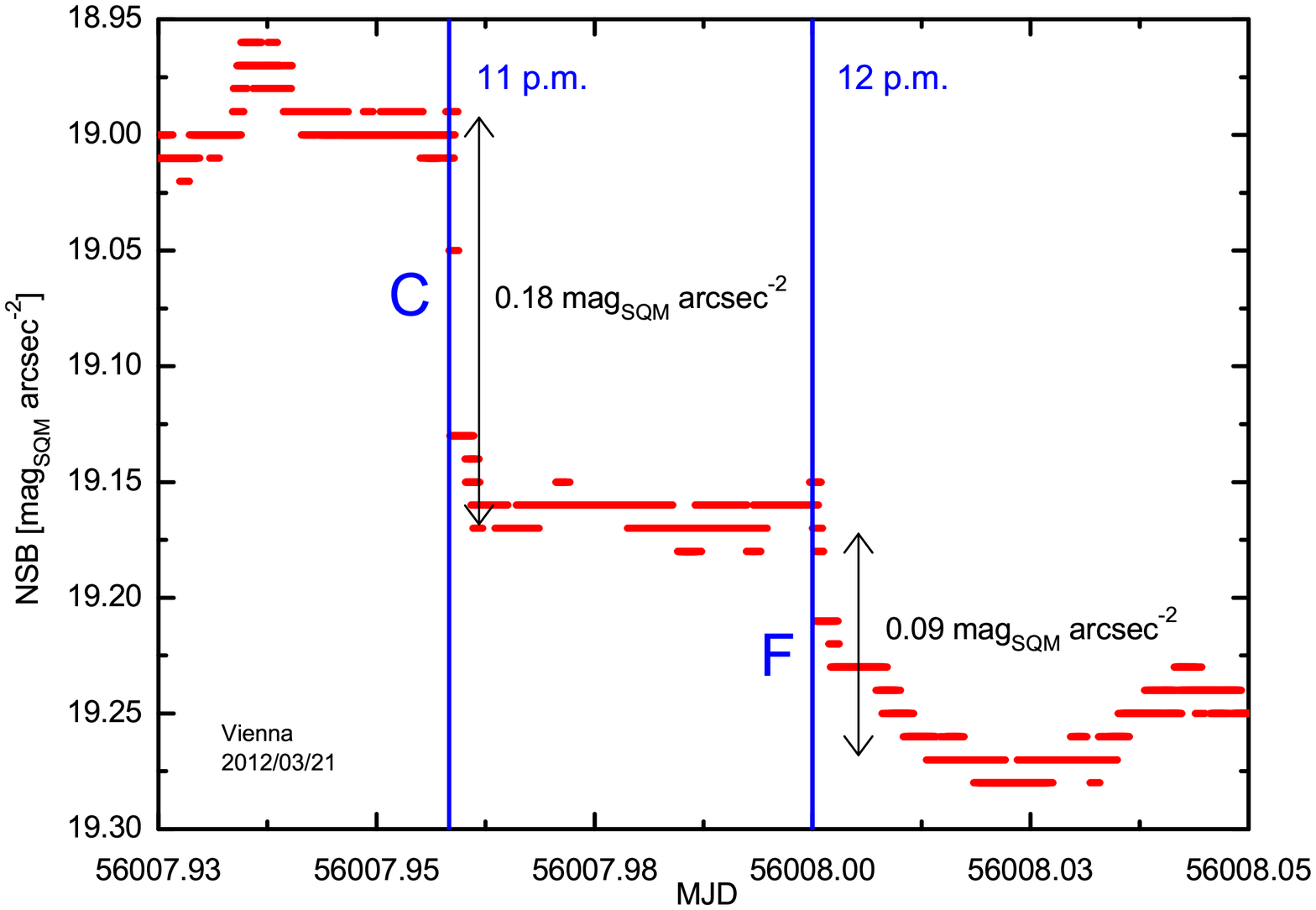}
	\caption{Detection of the curfew at 11 p.m. (C) and reduction of facade lighting at 12 p.m. (F) in Vienna on March 21st, 2012.}
	\label{fig:curfew}
\end{figure}

The city of Vienna reduces a substantial fraction of its street lighting at 11 p.m.\ (curfew, in the following: `C'). Main roads, however, are still kept at constant illuminance throughout the whole night.
At 12 p.m., most of the decorative facade lighting is switched off (this effect is called `F' below).
The sky brightness, therefore, shows two discontinuous steps towards lower values at 11 and 12 p.m., of which the first one is larger. This is shown in 
Fig.\ \ref{fig:curfew}.

On average, step C results in a night sky brightness decrease by 0.18 \magn, while step F leads to a decrease (= improvement of the sky quality) by 0.09 \magn.

It would be worthwhile to monitor similar effects of curfews in other cities and communities, also to demonstrate this benefit for the nocturnal environment.

\subsection{Influence of snow and other seasonal variations}
Snow has a strong influence on the NSB, especially in urban regions. The reflectance of snow is clearly much larger than that of bitumen. Hence, streets, pavements and squares covered by (fresh) snow reflect a huge portion of the downward directed radiation of streetlamps back to the sky. Moreover, a night with snowfall is per se a cloudy night -- hence, there is a second reason for such nights to be brighter, which is the backscattering of the upward light by the clouds. The combination of those two effects leads to extremely large values of the NSB during some winter nights.
The fact that the mean NSB values shown in the upper half of Fig. \ref{fig:lightcurve} reach values of 15\,\magn\ is due to the influence of clouded, snowy nights during the first weeks of 2013. This value corresponds to a zenithal luminance of 0.1\,\cdm, which is 461 times the natural NSB. Under the assumption of a uniform sky brightness of 0.1 \cdm\ (from the zenith to the horizon), an illuminance meter would display 0.31\,lx -- which is more than in the case of full moon. In reality, the sky is always brighter near the horizon than at the zenith, such that our measurement of 0.1\,\cdm, according to our experiences, corresponds to about 0.5\,lx. 

\begin{figure}
	\centering
	\includegraphics[width=10.5cm]{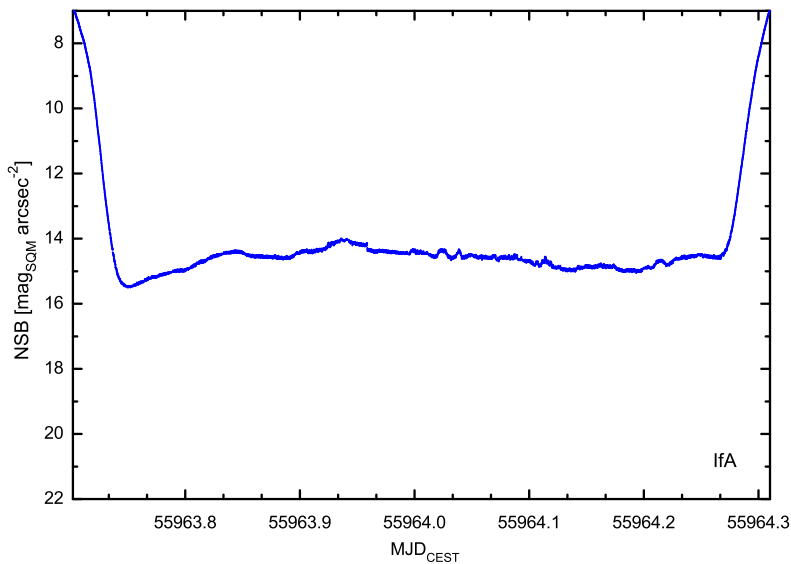}
	\caption{During the night from February 6th to February 7th 2012, the NSB reached a peak value corresponding to 0.27\,\cdm\ or 1570 times the natural zenithal sky brightness.}
	\label{fig:IFArecord}
\end{figure}

A night sky brightness of 15\,\magn\ or 0.1\,\cdm\ indicates very bright skies from the astronomical point of view, and yet we detected still larger NSB values at certain times -- not as mean values for whole nights, but as records lasting at least for some minutes to hours. For example, during the night from the 6th to the 7th of February 2012 (see Fig.\ \ref{fig:IFArecord}), up to 13.99\,\magn\ were recorded at 10:30 p.m. This spectacularly bright value corresponds to a luminance of 0.27\,\cdm\ or 1570 times the natural zenithal sky brightness. In this case, which typically occurs when there is snow cover and ongoing snow fall, the diffuse backscattered light from the sky alone would illuminate a landscape with more than 1\,lx. Under such conditions, one could even consider dimming down or switching off local illuminations of small streets and paths, since any landscape which is uniformly illuminated at 1\,lx appears brighter than the same landscape with ``punctual'' illuminations even at peak values of several lux.

\subsection{Special cases: thunderstorms, New Year's Eve}

\begin{figure}
	\centering
	\includegraphics[width=10.5cm]{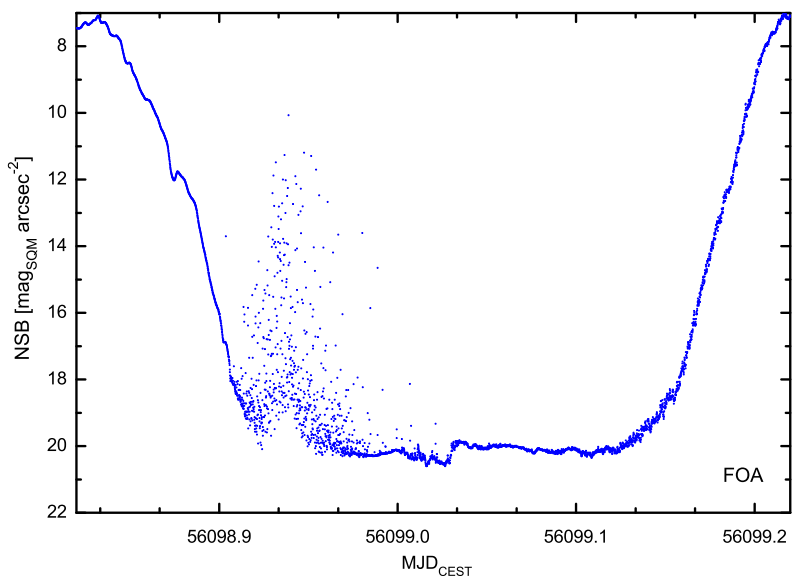}
	\caption{During the night from June 20th to June 21st 2012, a short term increase in the NSB from about 20 up to 10 \magn\ is found at our mountaintop observing station LFOA -- corresponding to an increase in the NSB by a factor of 10000 due to lightning.}
	\label{fig:FOAlightning}
\end{figure}

Our NSB measurements not only show the strong influence of lighting, but also the strong (but short-term) influence of lightning. Given that thunderstorms occur much more frequently during the summer months in Austria, the detection rate of lightning is highest between June and September.
Even with our sampling rate of one measurement every 7 seconds, bright lightning can have a strong impact in our night sky photometry. During two nights in June and August 2012, our NSB curves measured at Vienna (IfA) show ``spikes'' due to lightning with a baseline value of 16\,\magn\ and peak values up to 11\,\magn, which corresponds to an increase by a factor of 100. At the LFOA site, in contrast, we find even an increase by a factor of 10000 (!) due to lightning during the night from June 20 to June 21, 2012 (see Fig.\ \ref{fig:FOAlightning}).
Since lightning has extremely high luminance, but cover only a tiny fraction of the sky, it might be more interesting to measure the illuminances (in lux) caused by lightning. To the best of our knowledge, this has not been done in a systematic way so far.

A similar effect as the one caused by thunderstorms is observed -- at least in Vienna -- during New Year's Eve. The effect of local fireworks is again clearly detected in the NSB measurements. In this case, we measured a short-term increase of the sky brightness from 18 \magn\ to 16  \magn\ around midnight. Again, this is only a lower limit of the influence of fireworks on the NSB, since our SQMs are pointed toward the zenith, and hardly any fireworks rocket would reach the local zenith at the observatory, which is surrounded by a large park.

\subsection{Implications of our results for the transition from mesopic to scoptopic vision}

\begin{figure}
	\centering
	\includegraphics[width=11.5cm]{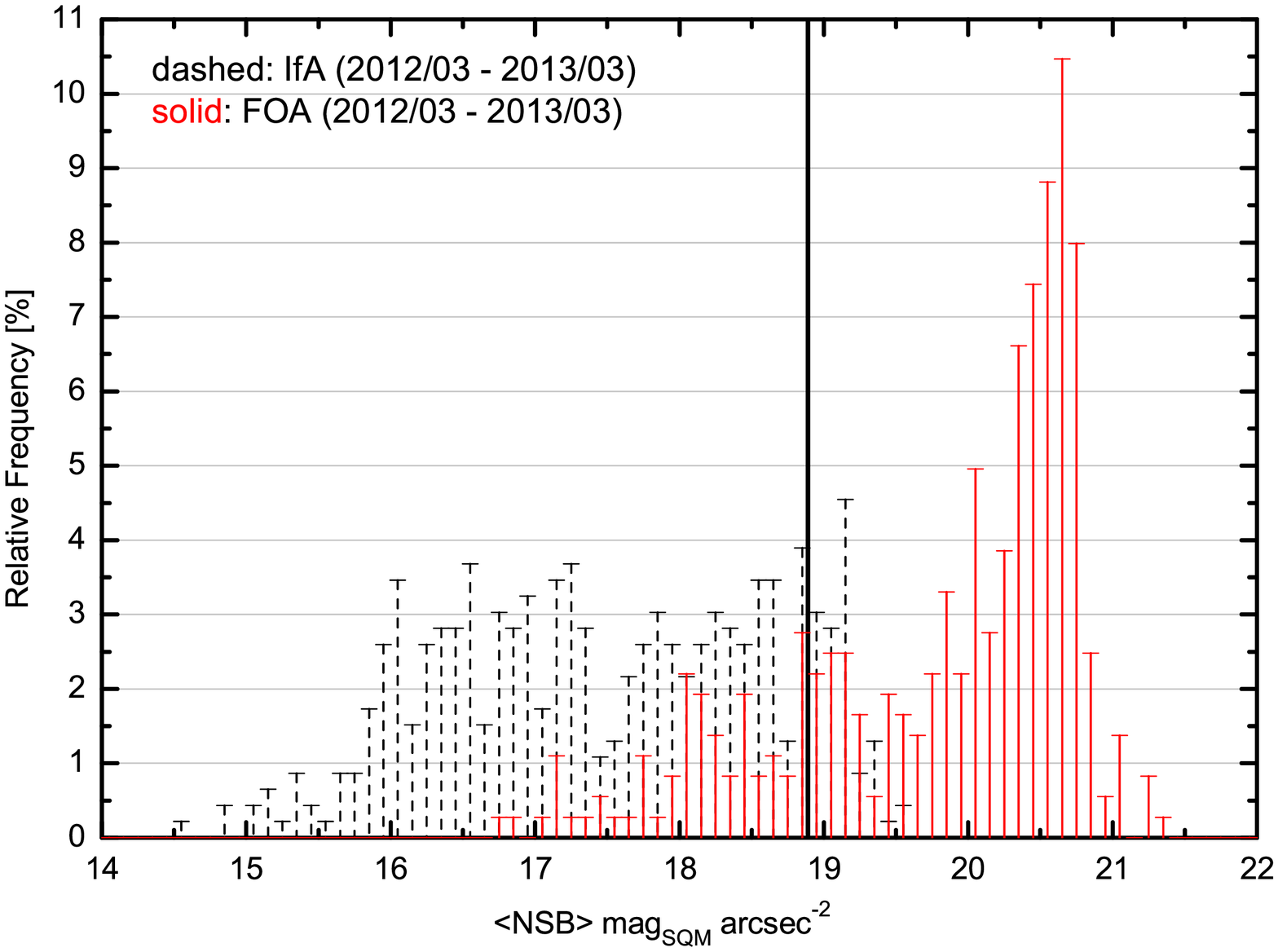}
	\includegraphics[width=11.5cm]{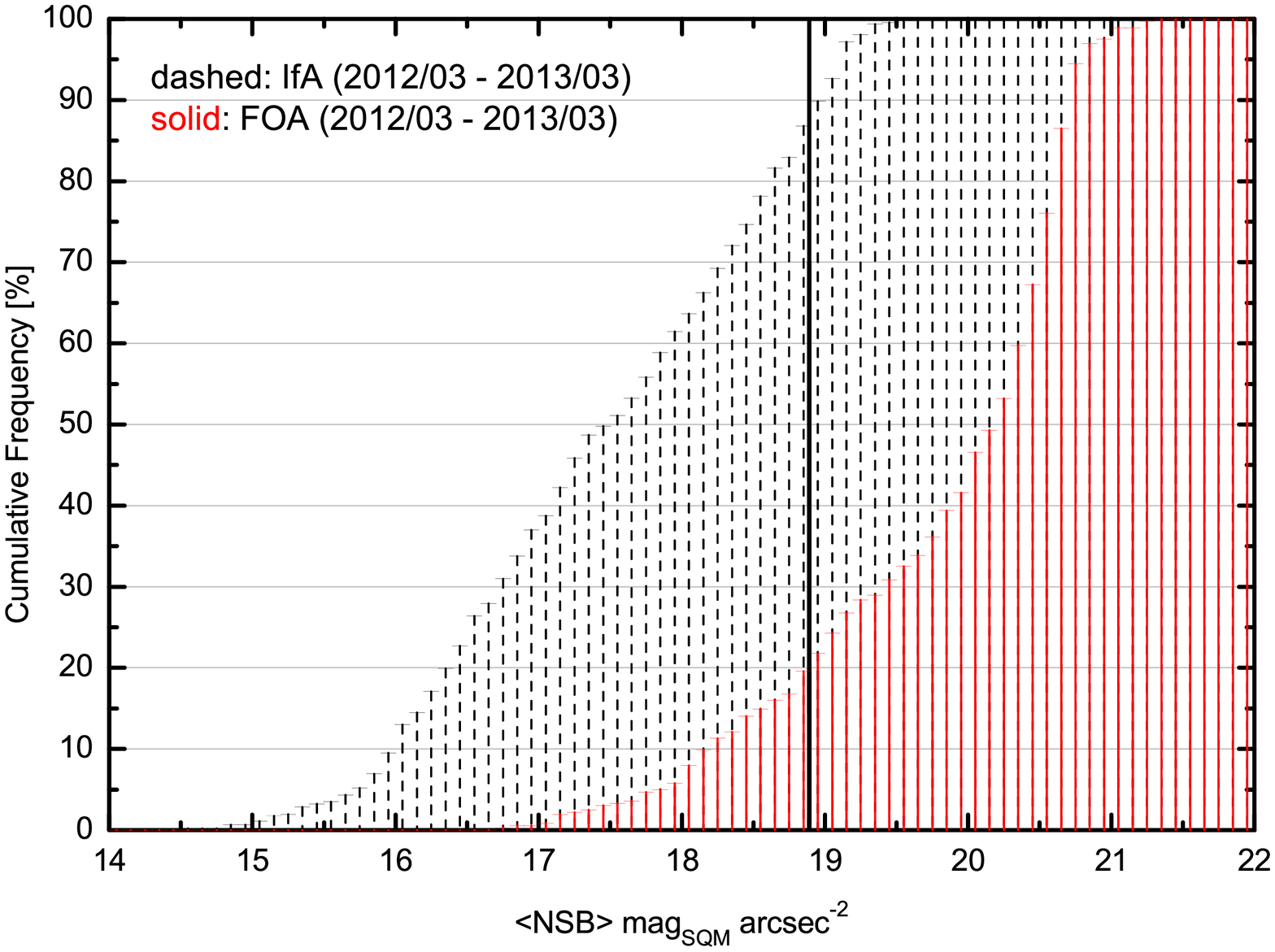}
	\caption{Histogram showing the relative (top) and cumulative (bottom) distribution of the mean NSB measured at IfA in Vienna (black dashed bars) and LFOA (red solid bars). The vertical line indicates the limit between scotopic and mesopic vision at \scotopiclimitmag.}
	\label{fig:histo}
\end{figure}

As shown in Figure \ref{fig:histo}, the mean zenithal NSB values in Vienna
(location IfA) are in the range 15\,\magn\ to
19.5\,\magn. The black vertical line represents a sky
brightness of \scotopiclimitmag. This corresponds to a
luminance of \scotopiclimitcdm or 3 \mcdm. We take this
value as the lower limit of the cones in human vision. At smaller
luminances, the color-insensitive rods entirely dominate our vision. If
the NSB is brighter than \scotopiclimitmag, the adaptation
of the human eye to darkness is very incomplete. The partial visibility of
our Milky Way at least near the zenith requires a value which is reported
to be close to 18.9 \magn\ as well. According to Fig.\ \ref{fig:histo}, almost 90\% of all nights at IfA are characterized by mean zenithal
NSBs larger than the ``\scotopiclimitmag'' threshold: during the vast majority of
all nights, therefore, entirely rod-dominated vision as well as Milky Way
visibility is not reached.

At our rural, 880m-above-sea-level observing station LFOA, in contrast,
only 20\% of all nights have mean zenithal NSB values brighter than 18.9
\magn. The full range of mean NSBs at LFOA is 17
\magn\  to 21 \magn. Note that
even the latter value is still about 1 magnitude (or a factor of 2.5)
above the natural zenithal night sky brightness.

\subsection{Spectral lines detected in the Night Sky}

\begin{figure}[t]
	\centering
	\includegraphics[width=13cm]{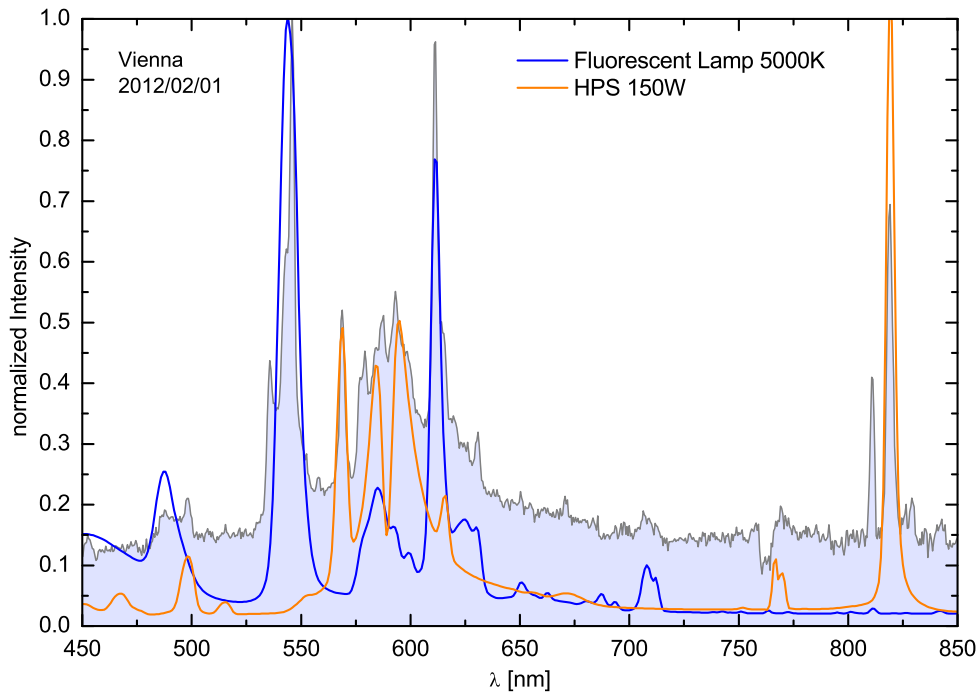}
	\caption{
		Night sky spectrum (filled area) obtained in Vienna on February 1st, 2012 towards SE. It can be clearly seen that the most prominent features correlate with emission lines originating from FL (blue solid line) and HPS (orange solid line). Laboratory spectra of street lamps can be found on the website of the National Oceanic and Atmospheric Administration \cite{NOAA}.}
	\label{fig:skyspectrum}
\end{figure}

Figure \ref{fig:skyspectrum} shows one of our night sky spectra, taken on
the 1st of February 2012. For comparison also laboratory spectra of common street lamps are shown. The strongest spectral line that we detect in the 
light-polluted night sky over Vienna is located at 546\,nm. An adjacent,
but weaker line is detected at 544\,nm.
Both lines are due to fluorescent lamps (FL), which are currently the most common type of lighting, especially in residential areas. 

The second strongest spectral feature that we find in our night sky spectra is centered at 611\,nm. Its origin is a composite of fluorescent lamps and high pressure sodium (HPS) lamps, which are used to illuminate suburban highways, crossroads and main streets. Between these two strongest spectral peaks, a series of lines are detected, most of which can be identified as sodium lines 
(outstanding peaks at 569\,nm, 588\,nm and 593\,nm). Another sodium line is
detected in the photographic infrared, at 819\,nm.
Since low-pressure sodium lines are uncommon in and around Vienna, the sodium peaks must be due to scattered light from high pressure sodium (HPS) lamps. More detailed information on the emission lines of HPS lamps can be found in Elvidge et al. \cite{Elvidge2010}.

Furthermore, atmospheric absorption bands can be identified at 762\,nm (A band) and 688\,nm (B band) due to magnetic dipole electronic transitions of the molecule O$_2$ (\cite{Kiehl1985}).

\begin{table}[t]
\caption{Selected (strong) spectral lines detected in Vienna's night sky.}
\begin{center}
\begin{tabular}{| c | c | c |}
  \hline
	Peak position [nm] & Assignment (see text) & Remark \\ \hline
	544 & FL & \\
	546 & FL & \\
	569/588/598 & HPS & \\
	611 & FL & blend with HPS \\
	819 & HPS & \\ \hline
\end{tabular}
\end{center}
\label{t:spectrallines}
\end{table}

According to the official statistics of the city of Vienna, 80\% of all
the lamps used for public illumination are fluorescent lamps, 16\% are
HPS, while other types of lamps add up to only 4\% by number. This is in good accordance with our spectroscopic results.

The total power used for street lighting in Vienna amounts to 14 MW
(value from the official website \url{http://wien.gv.at} \cite{WIEN}.
Assuming a lumen-to-Watt ratio of 70 for fluorescent lamps and 120
for HPS, we end up with a contribution of 784 Megalumen from the dominant
source (= fluorescent lamps) and 269 Megalumen from HPS lamps.
Together with the other 4\% (among them high pressure mercury vapor
lamps), the total light output of the city of Vienna, as far as it is caused
by street lighting, amounts to about 1.1 Gigalumen, or about 622 lumen per
capita.



\section{Conclusions}

After one year of NSB measurements at an urban site (in Vienna) and at a rural mountaintop, we may draw the following conclusions:

\begin{enumerate}[i]		


\item Due to its high sensitivity covering the bulk of the visible spectrum, Sky Quality Meters are appropriate not only for long-term studies of the NSB, but also for short-term events. Nevertheless, an adequate site, weather proof housing and shade are crucial, such that the sensor is protected against scattered (moon) light. Once tested, the instrument will monitor the NSB development within a relative error of approximately 5 \%.

\item The factor which is essentially determining the night sky brightness at light-polluted urban sites is the degree of cloudiness (see also Kyba et al. \cite{Kyba2012a}), since clouds strongly enhance the backscattering of artificial light in the atmosphere.

\item Typical NSB values at our urban measurement site IfA are 16.3 \magn\ or 33 \mcdm\ under overcast conditions. Note that without any artificial light, the overcast sky would be much darker than clear skies . The opposite is the case for cities nowadays.

\item Maximum values of the NSB at our urban measurement location are close to 14 \magn\ or 0.3 \cdm. This value refers exclusively to the diffuse (scattered) light from the night sky. It does not include any influence of lights shining directly into our detector. 

\item At our rural mountaintop site, the range of NSB values is typically 17 $\dots$ 21 \magn. Here, the phase of the moon remains the determining factor for the NSB, which probably implies a strong zeitgeber function of the moonlight at this rural place.

\item In contrast, the circalunar illuminance pattern is extinguished in urban and, very likely, to a large extent also in suburban regions. This will also influence the chronobiological rhythms of animals living in local ``islands of darkness'' such as big parks.

\item By a spectroscopic analysis of the light scattered back from the sky, we were able to identify lines from high pressure sodium lamps (e.g.\ at 569\,nm, at 588\,nm and at 593\,nm) and from fluorescent lamps (e.g.\ at 546\,nm and at 611\,nm) as main ``pollutants'' of the night sky.

\item For studying the night sky color distribution, filter systems e.g.\ Johnson UBV, should be used in the future. As indicated by our spectroscopic analysis, currently the blue end of the spectrum below 525nm does not show significant contribution to the NSB. This could change in the future, if lamps of high brightness temperatures will be used in the future (which would be unfortunate for many reasons: e.g.\ insect attraction, melatonin suppression, and enhanced light pollution).

\end{enumerate}


\begin{small}

{\noindent \bf Acknowledgements}

\noindent We thank Otto Beck, Erich Sch\"afer and Werner Zeilinger
for their help with the installation of the SQMs in Vienna
and especially at the Mittersch\"opfl Observatory.
We gratefully acknowledge Stefan Meingast's help with the spectroscopic measurements.
S. Uttenthaler acknowledges support from the Austrian Science Fund (FWF) under
project P 22911-N16.

\end{small}




\bibliographystyle{jqsrt}
\bibliography{bibliography}

\begin{thebibliography}{10}
\providecommand{\natexlab}[1]{#1}
\providecommand{\url}[1]{\texttt{#1}}
\providecommand{\href}[2]{#2}
\providecommand{\path}[1]{#1}
\providecommand{\eprint}[1]{\href{http://arxiv.org/abs/#1}{\path{#1}}}
\providecommand{\DOIprefix}{doi:}
\providecommand{\ArXivprefix}{arXiv:}
\providecommand{\URLprefix}{URL: }
\providecommand{\Pubmedprefix}{pmid:}
\providecommand{\doi}[1]{\href{http://dx.doi.org/#1}{\path{#1}}}
\providecommand{\Pubmed}[1]{\href{pmid:#1}{\path{#1}}}
\providecommand{\BIBand}{and}
\providecommand{\bibinfo}[2]{#2}
\ifx\xfnm\undefined \def\xfnm[#1]{\unskip,\space#1}\fi
\bibitem[{{Posch} et~al.(2010){Posch}, {Freyhoff} and {Uhlmann}}]{Posch2010}
\bibinfo{author}{{Posch}\xfnm[ T.]}, \bibinfo{author}{{Freyhoff}\xfnm[ A.]},
  \bibinfo{author}{{Uhlmann}\xfnm[ T.]}.
\newblock \bibinfo{title}{Das Ende der Nacht (German Edition)}.
\newblock \bibinfo{publisher}{Wiley-VCH}; \bibinfo{year}{2010}.
\newblock ISBN \bibinfo{isbn}{3527409467}.
\bibitem[{{Patat}(2008)}]{Patat2008}
\bibinfo{author}{{Patat}\xfnm[ F.]}.
\newblock \bibinfo{title}{{The dancing sky: 6 years of night-sky observations
  at Cerro Paranal}}.
\newblock \bibinfo{journal}{Astronomy and Astrophysics}
  \bibinfo{year}{2008};\bibinfo{volume}{481}:\bibinfo{pages}{575--591}.
\bibitem[{{Cinzano}(2005)}]{Cinzano2005}
\bibinfo{author}{{Cinzano}\xfnm[ P.]}.
\newblock \bibinfo{title}{{Night Sky Photometry with Sky Quality Meter}}.
\newblock \bibinfo{type}{Internal Report n. 9, v.1.4}; {Istituto di Scienza e
  Tecnologia dell'Inquinamento Luminoso (ISTIL)}; \bibinfo{year}{2005}.
\bibitem[{{Dobos}(2004)}]{Dobos2004}
\bibinfo{author}{{Dobos}\xfnm[ L.]}.
\newblock \bibinfo{title}{{John Hopkins University Filter Profile Services for
  the Virtual Observatory}}.
\newblock \bibinfo{howpublished}{Internet}; \bibinfo{year}{2004}.
\newblock \URLprefix \url{http://voservices.net/filter}.
\bibitem[{{Kyba} et~al.(2012){Kyba}, {Ruhtz}, {Fischer} and
  {H{\"o}lker}}]{Kyba2012a}
\bibinfo{author}{{Kyba}\xfnm[ C.C.M.]}, \bibinfo{author}{{Ruhtz}\xfnm[ T.]},
  \bibinfo{author}{{Fischer}\xfnm[ J.]}, \bibinfo{author}{{H{\"o}lker}\xfnm[
  F.]}.
\newblock \bibinfo{title}{{Red is the new black: how the colour of urban
  skyglow varies with cloud cover}}.
\newblock \bibinfo{journal}{Monthly Notices of the RAS}
  \bibinfo{year}{2012};\bibinfo{volume}{425}:\bibinfo{pages}{701--708}.
\bibitem[{{Cinzano} et~al.(2001){Cinzano}, {Falchi} and
  {Elvidge}}]{Cinzano2001}
\bibinfo{author}{{Cinzano}\xfnm[ P.]}, \bibinfo{author}{{Falchi}\xfnm[ F.]},
  \bibinfo{author}{{Elvidge}\xfnm[ C.D.]}.
\newblock \bibinfo{title}{{The first World Atlas of the artificial night sky
  brightness}}.
\newblock \bibinfo{journal}{Monthly Notices of the RAS}
  \bibinfo{year}{2001};\bibinfo{volume}{328}:\bibinfo{pages}{689--707}.
\bibitem[{Elvidge et~al.(2010)Elvidge, Keith, Tuttle and Baugh}]{Elvidge2010}
\bibinfo{author}{Elvidge\xfnm[ C.D.]}, \bibinfo{author}{Keith\xfnm[ D.M.]},
  \bibinfo{author}{Tuttle\xfnm[ B.T.]}, \bibinfo{author}{Baugh\xfnm[ K.E.]}.
\newblock \bibinfo{title}{Spectral identification of lighting type and
  character}.
\newblock \bibinfo{journal}{Sensors}
  \bibinfo{year}{2010};\bibinfo{volume}{10}(\bibinfo{number}{4}):\bibinfo{pages}{3961--3988}.
\bibitem[{{Kiehl} and {Yamanouchi}(1985)}]{Kiehl1985}
\bibinfo{author}{{Kiehl}\xfnm[ J.T.]}, \bibinfo{author}{{Yamanouchi}\xfnm[
  T.]}.
\newblock \bibinfo{title}{{A parameterization for absorption due to the A, B,
  and gamma oxygen bands}}.
\newblock \bibinfo{journal}{Tellus Series B Chemical and Physical Meteorology
  B} \bibinfo{year}{1985};\bibinfo{volume}{37}:\bibinfo{pages}{1--6}.
\bibitem[{{US Department of Commerce, NOAA, National Geophysics Data
  Center}(2013)}]{NOAA}
\bibinfo{author}{{US Department of Commerce, NOAA, National Geophysics Data
  Center}\xfnm[]}.
\newblock \bibinfo{title}{{Laboratory Spectra}}.
\newblock \bibinfo{howpublished}{Internet}; \bibinfo{year}{2013}.
\newblock \URLprefix \url{http://www.ngdc.noaa.gov/dmsp/spectra.html}.
\bibitem[{{Wien Leuchtet (Magistratsabteilung 33)}(2013)}]{WIEN}
\bibinfo{author}{{Wien Leuchtet (Magistratsabteilung 33)}\xfnm[]}.
\newblock \bibinfo{title}{{Wiens Beleuchtung in Zahlen}}.
\newblock \bibinfo{howpublished}{Internet}; \bibinfo{year}{2013}.
\newblock \URLprefix
  \url{http://www.wien.gv.at/verkehr/licht/beleuchtung /oeffentlich/zahlen.html}.

\end{thebibliography}








\appendix

\section{Discussion on the origin of the scatter while parallel measurements of two SQMs}

\begin{figure*}[t]
	\centering
	\includegraphics[width=13cm]{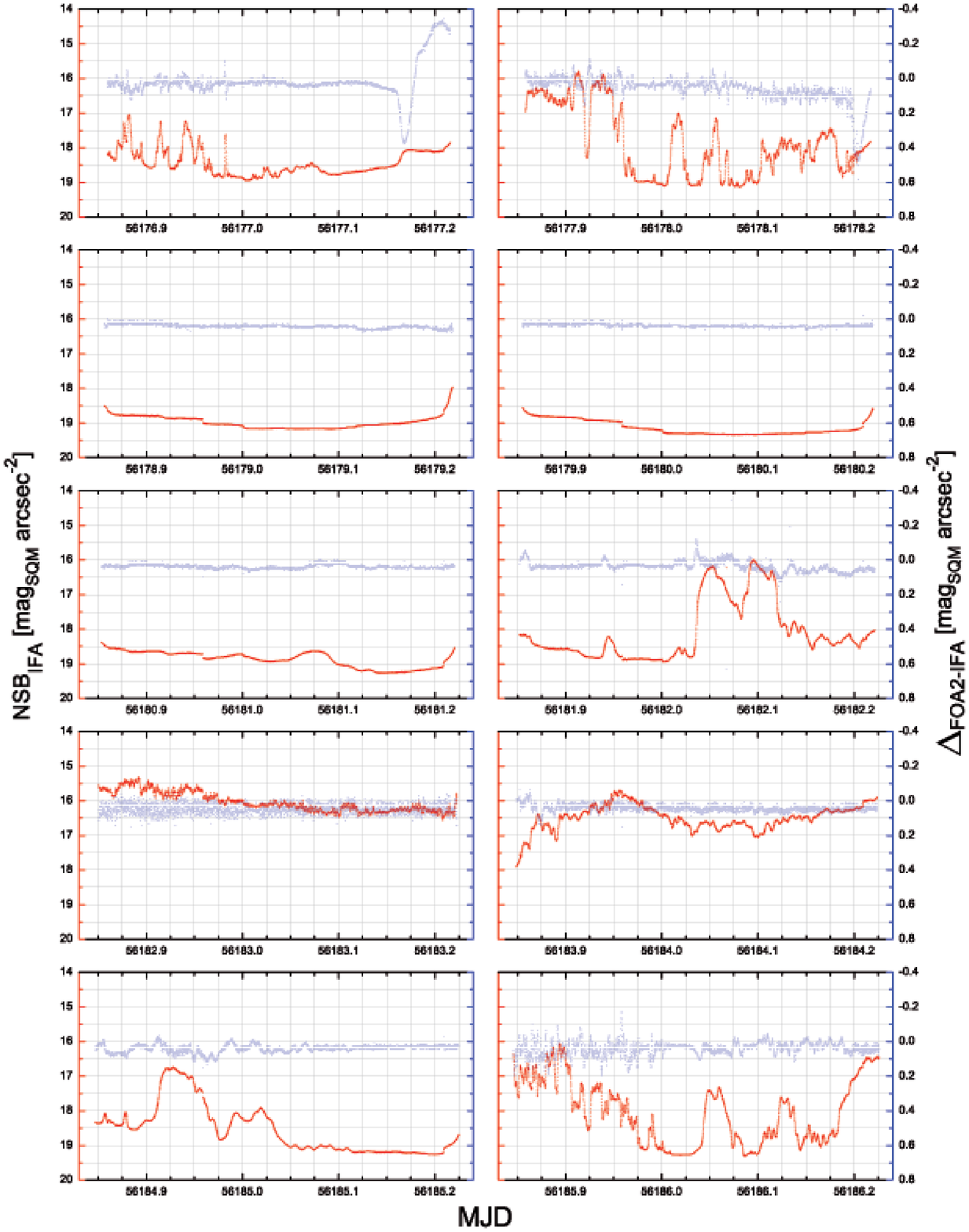}
	\caption{Intercomparison of our devices from September 6th to 16th, 2012. The scale on the left indicates the measured NSB in \magn\ (red curve) for one of the devices, whereas the scale on the right shows the difference in \magn\ (blue curve) of the measured values between the two parallel mounted devices.}
	\label{fig:qualitylightcurves1}
\end{figure*}

\begin{figure*}[t]
	\centering
	\includegraphics[width=13cm]{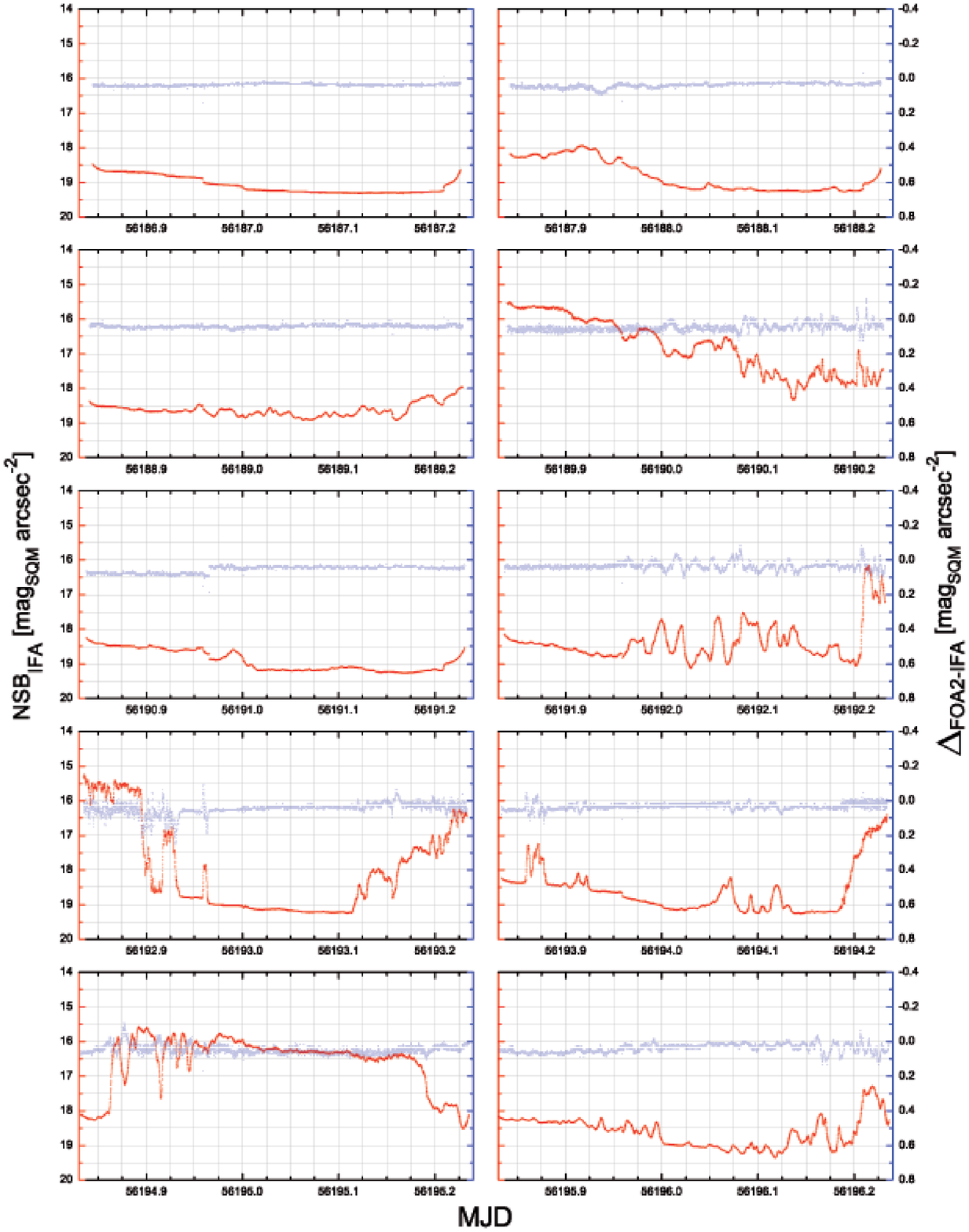}
	\caption{Intercomparison of our devices from September 16th to 26th, 2012.  The scale on the left indicates the measured NSB in \magn\ (red curve) for one of the devices, whereas the scale on the right shows the difference in \magn\ (blue curve) of the measured values between the two parallel mounted devices.}
	\label{fig:qualitylightcurves2}
\end{figure*}

\begin{figure*}[t]
	\centering
	\includegraphics[width=13cm]{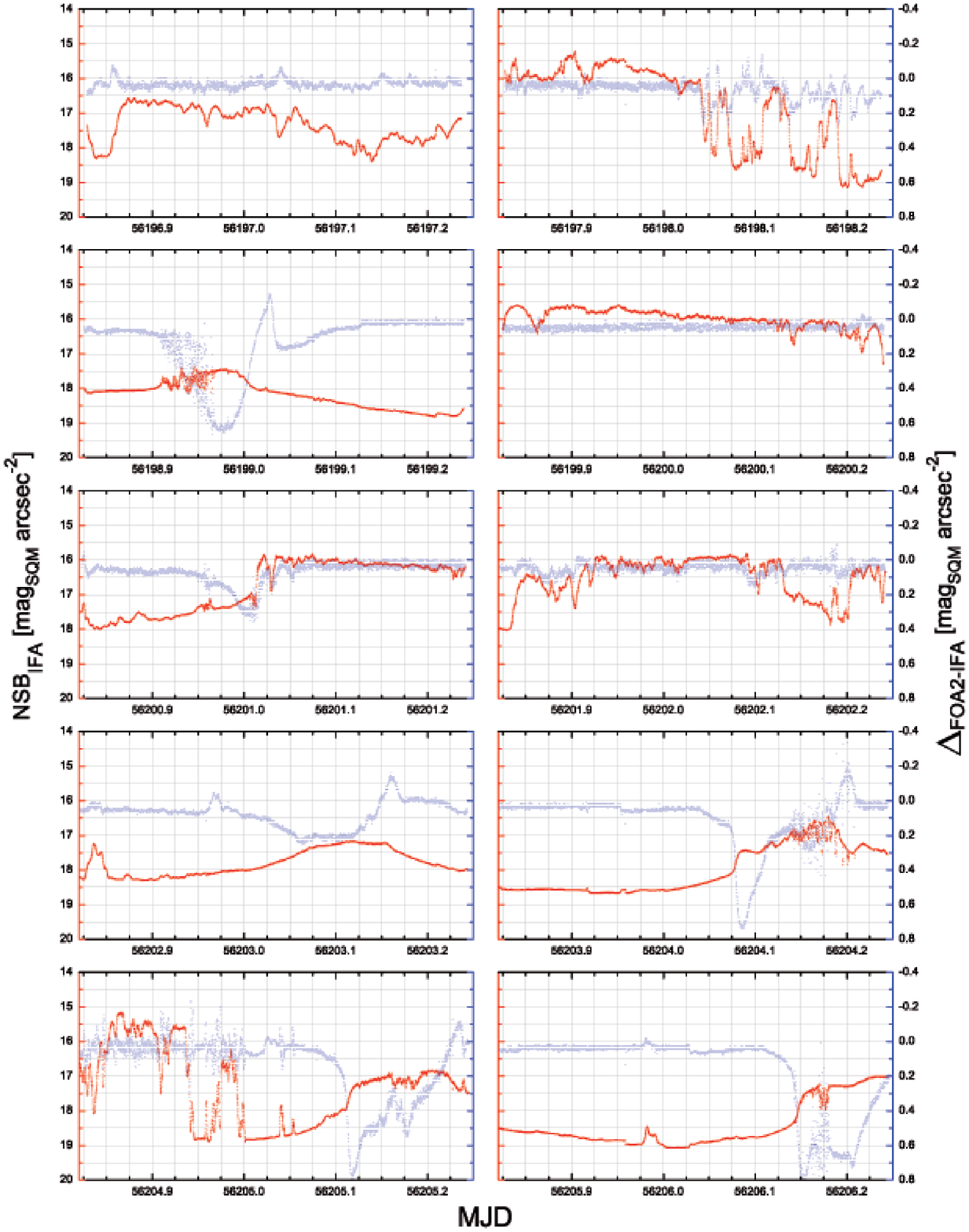}
	\caption{Intercomparison of our devices from September 26th to October 6th, 2012.  The scale on the left indicates the measured NSB in \magn\ (red curve) for one of the devices, whereas the scale on the right shows the difference in \magn\ (blue curve) of the measured values between the two parallel mounted devices.}
	\label{fig:qualitylightcurves3}
\end{figure*}

We performed parallel measurements of our two SQM devices over a period of 33 days. The daily light curves showing the measured NSB value and the difference between the readings of the devices are shown in Figs. \ref{fig:qualitylightcurves1} to \ref{fig:qualitylightcurves3}. 

\begin{figure}
	\centering
	\includegraphics[width=9cm]{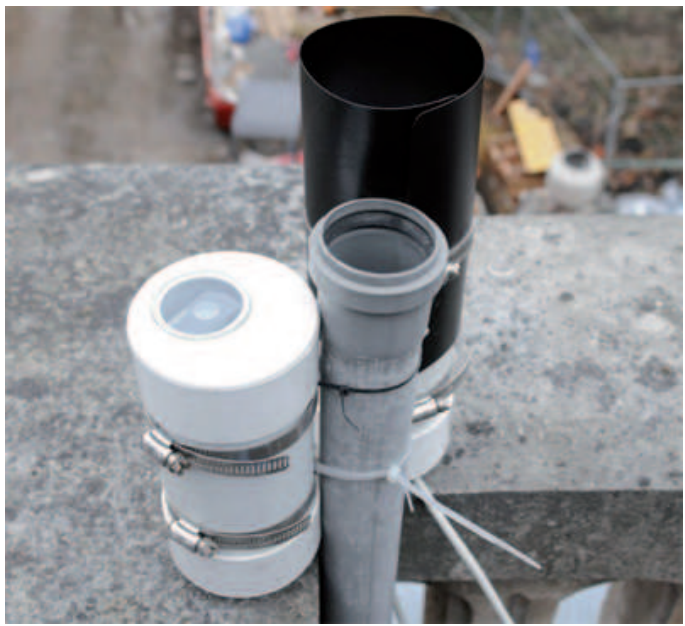}
	\caption{Parallel measurements of two SQMs. One of the devices is equipped with a simple shade for preventing scattered moon light.}
	\label{fig:sqms}
\end{figure}

It can be seen that both devices gave the same or similar response for most of the measurements and the calibration of the devices as delivered by the manufacturer is very accurate. However, within certain time intervals the differences increased up to 0.9\,\magn. After checking for moon phase, altitude and azimuth, it was realized that these big differences always occur 1) at certain alt/azimuth coordinates of the moon, 2) when the moon illuminated fraction was greater than 50 \% and 3) in the same ``sine-wave-like'' shape. Thus it is clear that the differences are a result of scattered moonlight. We assume that this affects only the SQM-LE, the lensed SQM version and suggest to use some sort of shielding or shading (see Fig.\ \ref{fig:sqms}) when using the device. Since the lensed version has a very narrow field of view of only $\approx$20$^{\circ}$, some sort of shade can be installed easily.

\end{document}